\documentclass{aa}  
\usepackage{graphicx}
\usepackage{amssymb}
\usepackage{txfonts}
\usepackage{mathrsfs}
\usepackage{array}
\usepackage{natbib}
\bibpunct{(}{)}{;}{a}{}{,}   
\begin{document}
   \title{Alpha element abundances and gradients in the Milky Way bulge from FLAMES-GIRAFFE spectra of 650 K giants\thanks{Based on observations collected at the European Southern Observatory, Paranal, Chile (ESO programmes 071.B-0617 and 073.B-0074}}
   
   \author{O. A. Gonzalez$^{1}$  \and M. Rejkuba$^{1}$ \and M.    Zoccali$^{2}$  
 \and V. Hill$^{3}$  \and G. Battaglia$^{1}$ \and C. Babusiaux$^{4}$ \and D. Minniti$^{1,2,8}$ \and B.
Barbuy$^{5}$ \and A. Alves-Brito$^{2}$ \and A. Renzini$^{6}$ \and A. Gomez$^{4}$ \and S. Ortolani$^{7}$}
   
   \offprints{O. A. Gonzalez}
   \institute{ $^{1}$European
   Southern Observatory, Karl-Schwarzschild-Strasse 2, D-85748 Garching,
Germany\\ \email{ogonzale@eso.org; mrejkuba@eso.org; gbattagl@eso.org}\\
   $^{2}$Departamento    Astronom\'ia    y Astrof\'isica,
   Pontificia Universidad  Cat\'olica de Chile,  Av. Vicu\~na Mackenna
   4860,         Stgo.,         Chile\\         \email{mzoccali@astro.puc.cl;
Dante@astro.puc.cl}\\
   $^{3}$Universit\'e de Nice Sophia Antipolis, CNRS, Observatoire de la
C\^{o}te d'Azur, B.P. 4229, 06304 Nice Cedex 4, France\\
\email{vanessa.hill@obspm.fr}\\
   $^{4}$Observatoire de Paris-Meudon, 92195   Meudon    Cedex,   France\\  
\email{Ana.Gomez@obspm.fr; carine.babusiaux@obspm.fr}\\ 
   $^{5}$Universidade de   S\~{a}o   Paulo,   IAG,   Rua  do   Mat\~{a}o   1226,
  Cidade
   Universit\'aria,     S\~{a}o      Paulo     05508-900,     Brazil\\
\email{Barbuy@astro.iag.usp.br}\\
   $^{6}$INAF-Osservatorio Astronomico di  Padova, Vicolo dell'Osservatorio 
2,I-35122 Padova, Italy\\  \email{alvio.renzini@oapd.inaf.it}\\
   $^{7}$Universit\`a di Padova,   Dipartimento 
di  Astronomia,   Vicolo  dell'Osservatorio 5,I-35122 Padova, Italy\\
\email{sergio.ortolani@unipd.it} \\   
   $^{8}$Specola Vaticana, V00120 Citta’ del Vaticano, Italy\\}

   \date{Received ; Accepted }

  
\abstract  
{}
{We present the  analysis of the [$\alpha$/Fe] abundance  ratios for a
large number of stars at several locations in the Milky Way bulge with the aim of constraining its
formation scenario.}
{We obtained FLAMES-GIRAFFE spectra (R=22,500) at the ESO Very Large Telescope for 650 bulge red giant branch (RGB) stars and performed spectral synthesis to measure Mg, Ca, Ti, and Si abundances.
This  sample is  composed  of  474 giant  stars 
observed in 3 fields along the minor axis of the Galactic bulge and at
latitudes b=-4$^\circ$,  b=-6$^\circ$, b=-12$^\circ$.  Another  176 
stars  belong to a field containing  the globular cluster
NGC 6553, located at b=-3$^\circ$ and 5$^\circ$ away from the other three fields 
along the major axis.   Stellar parameters and metallicities for  these stars were
presented in  Zoccali et al.\  (2008). We have also  re-derived stellar
parameters and  abundances for the sample  of thick and  thin disk red
giants  analyzed in  Alves-Brito  et al.\  (2010).   Therefore using  a
homogeneous abundance database for the  bulge, thick and thin disk, we
have performed a differential analysis minimizing systematic errors, to
compare the formation scenarios of these Galactic components.}
{Our  results  confirm, with  large  number  statistics, the  chemical
similarity between the Galactic bulge and thick disk, which are both enhanced in
alpha elements when compared to the thin disk. In the same context, we analyze
[$\alpha$/Fe] vs.  [Fe/H] trends across different bulge regions. 
The most metal rich stars, showing low [$\alpha$/Fe]  ratios  at b=-4$^\circ$  
disappear  at higher Galactic latitudes in  agreement with the observed 
metallicity  gradient in the bulge. Metal-poor stars  ([Fe/H]$<-0.2$) show a remarkable homogeneity
at different bulge locations.}
{We have obtained further constrains for the formation scenario of the
Galactic  bulge.  A metal-poor component chemically indistinguishable
from the thick disk hints for a fast and early formation for both the
bulge  and the  thick disk.  Such a  component shows  no  variation, neither in abundances nor kinematics,  among different  bulge  regions.  A  metal-rich   component   showing  low
[$\alpha$/Fe] similar to  those of the thin disk  disappears at larger
latitudes. This allows us to trace a component formed
through fast  early  mergers  (classical bulge)  and  a  disk/bar
component formed on a more extended timescale.}
             
\authorrunning{O.A. Gonzalez et al.}
\titlerunning{Galactic bulge $\alpha$-element abundances}

\maketitle

%

\section{Introduction}

Nearly 25\% of the  visible  light  emitted  from  stars  in  the  local
universe comes from bulges embedded in galactic disks
\citep{Fukugita98}.    Therefore,   a    fundamental   step   in   the
understanding  of  galaxy formation  and  evolution  is  to obtain  an
accurate characterization of these structures. The picture seems to be
complicated as recent observations suggest that bulges may be an in-homogeneous 
class of objects, each of  them holding a different formation  history \citep[e.g.\ ][]{
athana05}. In this context, The Milky Way bulge is a unique laboratory
in which we can study in detail the stellar populations as well as the
chemical   composition  and   kinematics  on   a   star-by-star  basis
\citep[see ][for a review]{minniti08}.

In our  galaxy, the  consensus is that  the bulk  of the bulge  is old
(t$\sim$12  Gyr) with  a  metallicity distribution  that extends  from
[Fe/H]=$-1.5$  to  [Fe/H]=$0.5$,  with  a peak  at  solar  metallicity
\citep[e.g.][and references  therein]{zoccali03,zoccali08, clarkson08, brown10} and
that the formation of the bulge  was fast as evidenced by the enhanced alpha 
elements in bulge stars \citep{mcwilliam94, rich05, zoccali06,fulbright06,fulbright07,lecureur07, melendez08, ryde09, ryde10}. 
A vertical metallicity gradient is also observed \citep{minniti95,zoccali08} although 
it appears to be absent in the inner bulge regions (b$<$4$^\circ$) \citep{rich_origlia07}. Therefore, the  conclusions  so  far  regarding  age,  metallicity  and  chemical
enrichment in the bulge point  to a formation scenario similar to that
of early-type  galaxies through one or more starburst  events triggered by
mergers in the early ages of the universe \citep{weinzirl09,hopkins10} -- in other words, 
as a {\it classical bulge}. 

On the other hand,   \citet{kormendy93}  presented observations of 
a large number of bulges with properties resembling more disk-like
structures rather than the  spheroidal concept of  classical bulges
\citep[see][for a review]{kormendy04}.   These   structures,  named
\textit{pseudo-bulges} or \textit{disky bulges}, tend to show younger
stellar  populations, to be  rotation supported,  and to  have surface
brightness profiles  similar to those  of disks and  less concentrated
than those of classical bulges. They  are expected to be formed from
gas inflows  driven by  disk instabilities such  as the presence  of a
central bar, which are common phenomena in spiral galaxies. Galaxies,  when  observed  edge-on, often  show  a
central component which swells out of the disk with a boxy, peanut, or
even an  X shape \citep[e.g.][]{bureau06,patsis02}.  Several numerical simulations have shown that  the inner parts of a bar might
evolve and buckle  off the plane of the  disk showing these particular
shapes \citep{athana05, debattista04, debattista06}.

The above scenarios for bulge formation have historically emerged
from the observation of bulges in nearby galaxies. However, a new
paradigm has recently emerged, mostly driven by the observations of
galaxies at redshift $\sim 2$, i.e., at a lookback time comparable to
the age of bulge and thick disk stars. Such galaxies may offer examples of how 
our own Milky Way was looking like in its early days, prior and during 
its bulge formation. At $z\sim 2$ a
widespread population exists of large, rotating disk galaxies, with
much higher gas fractions and velocity dispersion compared to local
spirals \citep{genzel06,genzel08, fs09, tacconi10, daddi10}.
Such gas-rich galaxies are
prone to disk instabilities that can result in bulge formation over
timescales of a few $10^8$ yr, much shorter than those typically
ascribed to secular instabilities in local disk galaxies \citep{immeli04, carollo07, elmegreen08}.
Such rapid formation, partly due to
the quasi-exponential mass growth experienced by these galaxies, quite
naturally leads to an $\alpha$-element enhancement in the resulting
stellar populations \citep{renzini09,peng10}, that would apply
both to the disk and bulge stars formed at these early cosmic epochs.

While the existence of  a metallicity gradient
in the  bulge, as shown  by \citet{zoccali08} strongly  argues against
the pseudo-bulge hypothesis of \citet{kormendy04}, some recent chemical 
abundance studies and morphological signatures have provided supporting 
evidence for such a scenario. Using a sample  of high resolution spectra for disk
and bulge  giants, \citet{melendez08}  find a similarity  between the
[O/Fe]  ratio of  the bulge  and the  thick disk,  both  enhanced when
compared to the thin disk. Later on, \citet{alves10} reach the same
conclusions when analyzing other  $\alpha$-elements from the sample of
\citet{melendez08}. These results are in disagreement with previous results 
which show a bulge more enhanced in alpha elements than the thick disk. 
Since those previous studies were done comparing bulge
giants to local disk dwarfs from other works \citep{bensby05,reddy06},
\citet{melendez08} claim that differences between thick disk and bulge
stars might be explained by systematic errors when comparing dwarfs to
giants or relative calibrations between those works. If the similarity
between the  bulge and thick disk  is confirmed with a larger and homogeneous sample,
this  would  allow  for  a   common  origin  for  these  two  galactic
components,  during the  early life  of  the Galaxy.   However, it  is
important to recall  that the origin of the thick disk is also unknown,
though several hypothesis have been explored \citep[][and references therein]{dimatteo11}, 
and most of the results are based on stars from the solar neighbourhood. 
Progress on this point has been achieved  for the
first time  by \citet{bensby10b}  who provided a  sample of  inner disk
stars  which show no  significant chemical  differences with respect to  local disk
giants.

In terms of morphology, the evidences for a barred structure in
the Milky Way are numerous. Red clump stars distributions from photometric
surveys have established the existence of a bar inclined with respect to the
Sun--Galactic center line of sight with its near end towards positive longitudes.
Observations of the bulge are usually modeled with a triaxial bulge with a
position angle of $\sim$15-30 degrees and nearly 2.5 kpc in length \citep[e.g.][]{babusiaux05,rattenbury07}. 
\citet{howard09}  analyzed the kinematics of the bulge concluding that the dynamical
signature of the stars at high latitudes (b=-8$^\circ$) is the same as that in fields
closer to the galactic plane (b=-4$^\circ$), therefore showing evidence for cylindrical
rotation {\it as expected for boxy bulges}. 
Their data lack evidence indicative of an accretion origin of bulge in contrast with the claims of tidal streams in kinematics and metallicity from \citet{rangwala09-2}. Moreover their cylindrical rotation signal
was modelled by \citet{shen10} without any need for a classical component.

There has also been evidence, from both observations and models, that different
types of bulges could coexist in the same galaxy 
\citep{samland03,nakasato03,kormendy04,peletier07}. If this is the case, we might have 
a very complex structure in the bulge, with two or more components partially overlapping.

Evidence for a dual nature of our bulge has most recently been invoked based on chemical abundances and kinematic studies. Indication for bimodality in the metallicity distribution has been
found from red clump stars in Baade's Window (Hill et al.\ 2011 submitted). The
two components seem to have also different kinematics \citep{babusiaux10}
which is consistent with different structures present in the bulge: 
a metal-rich with bar-like kinematics and a metal-poor one showing 
kinematics of a spheroidal component.

The picture is further complicated by the recent finding of a double peaked
red clump along the  minor axis, at both positive and negative
latitudes, but only for $\mid b \mid > 6^\circ$. The double clump is clearly visible in several  
color magnitude diagrams available in the literature  \citep[e.g.][]{zoccali03,rattenbury07,mcwilliam10,nataf10}. Such a feature has been interpreted 
as two components at different distances and is likely to be the outcome 
of an X-shaped bulge \citep[][]{mcwilliam10}.

All these observational evidences  have suggested a very complex structure of the 
bulge, with  two or more distinct components partially overlapping. Signatures
of the different formation histories for these components, should still remain
in bulge stars.

In this paper we analyze the $\alpha$-element distribution of giant stars in three fields along the minor axis and another at a larger galactic longitude. We use them to homogeneously compare with disk giant stars abundances and to investigate the presence of gradients within the bulge.

\section{The sample}

The bulge spectra discussed here belong to the same sample presented 
in \citet[][hereafter Z08]{zoccali08} as
part of a project aimed to provide a spectroscopic characterization of the
Galactic bulge. The sample consists of spectra of 650 K giant stars obtained with the
FLAMES-GIRAFFE spectrograph \citep{pasquini03} using the HR13, HR14 and HR15 setups (6100-7000\AA{}) at a resolution of R$\sim$ 20,000. For the present analysis we have only used the HR13 setup (R=22,500) in which we have lines for all the analyzed elements. As described in Z08, the S/N of our sample ranges from 40 to 90. Details on the fields analyzed here are presented in Table~\ref{i1}. In the last column of Table~\ref{i1} we list the number of stars for which we were able to measure abundances for Ca, Mg, Ti and Si. These are only those spectra that had sufficient S/N ($>$50) for all the elements to be measured and did not present very large FWHM which could affect the results due to line blending. The sample also contains target stars which belong to bulge globular clusters: NGC 6522 in Baade's Window (7 stars), NGC 6558 in the field at b=-6$^\circ$ (3 stars) and NGC 6553 in the eponymous field (29 stars). These stars were clasified as cluster members by Z08 based on the following criteria: [Fe/H] within 0.2 dex and radial velocity within $\pm 10$~km/s of the mean for the cluster, and a location within 2 arcmin from the cluster center.
 
\begin{figure}
\begin{center}
\includegraphics[scale=0.45,angle=0]{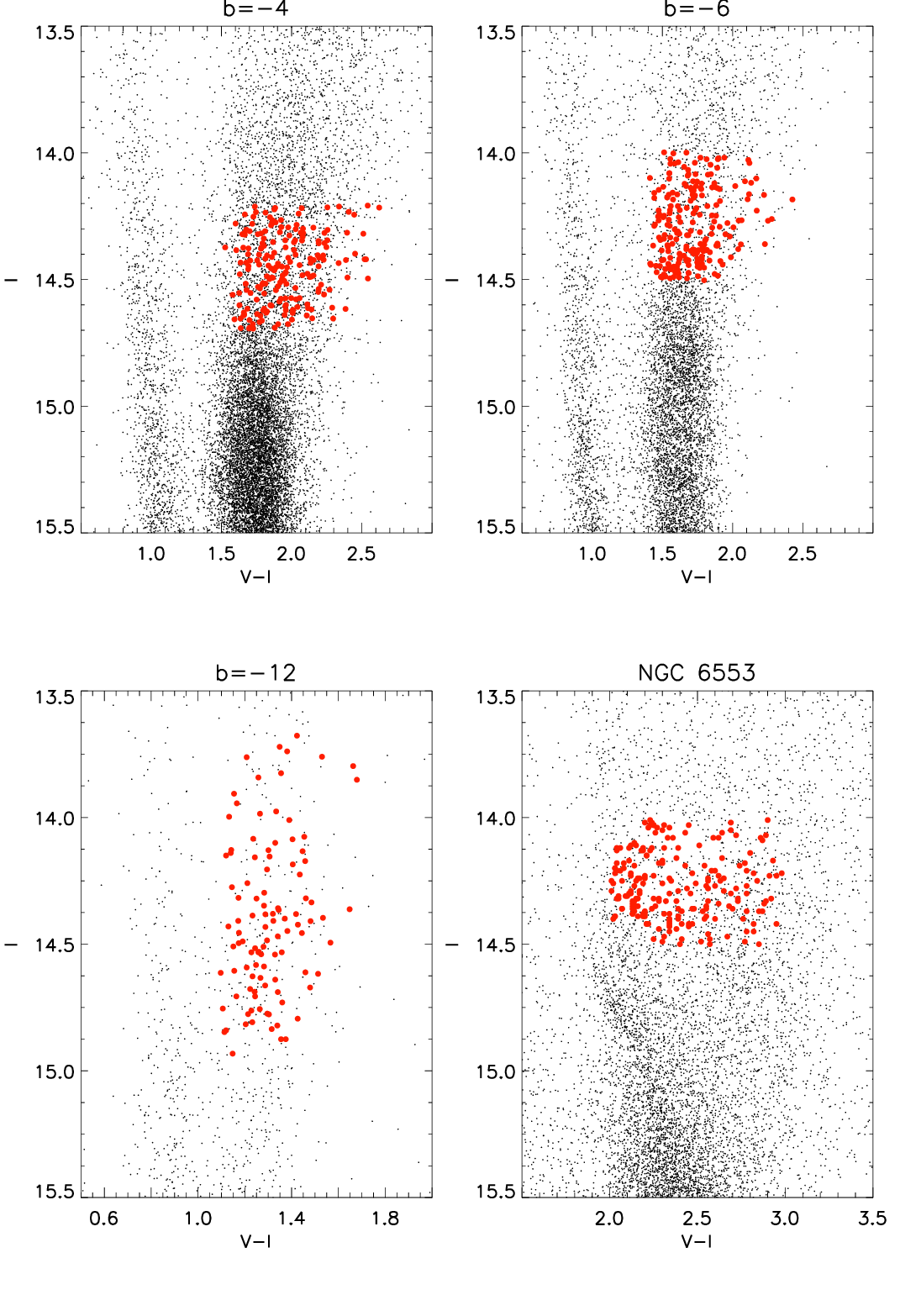}
\caption{Color  magnitude  diagrams  for the four fields: Baade's  window
(\textit{upper left}), b=-6$^\circ$ (\textit{upper right}), b=-12$^\circ$ (\textit{lower left}) and NGC6553 (\textit{lower right}). Spectroscopic targets in each field are marked as large filled circles. Magnitudes were obtained from the OGLE
catalogue \citep{udalski02} and the Z08 catalog obtained from WFI images}
\label{cmdfull}
\end{center}
\end{figure}

Magnitudes and astrometry for this sample come from the OGLE catalog for Baade's Window and from WFI images for the field at b=-6$^\circ$, b=-12$^\circ$ and NGC6553 (see Z08 for a detailed description). Selected targets are nearly 1 mag above the red clump as shown in Fig~\ref{cmdfull}.

\begin{table}
\begin{center}
\caption{Galactic coordinates of bulge fields. Extinction values and number of stars analyzed in each field are also listed. \label{i1}}
\begin{tabular}{c c c c c c}
\\[3pt]
\hline
N & Name & \textit{l$^\circ$} & \textit{b$^\circ$} & E(B-V) & stars\\ 
\hline
1 & Baade's Window & 1.14 & -4.18  & 0.55 & 194\\
2 & $b=-6^{o}$ & 0.21 & -6.02 & 0.48 & 194\\
3 & Blanco field & 0.00 & -12.0 & 0.20 & 86 \\
4 & NGC 6553 & 5.25 & -3.02 & 0.70 & 176 \\
\hline
\end{tabular}
\end{center}
\end{table}

The sample for thick and thin disk RGB stars in the local neighborhood 
consists of the same spectra analyzed in
\citet{melendez08} and \citet[][hereafter AB10]{alves10}. Stars belong to a range of
effective temperatures of 3800 $<$ T$_{eff}$ $<$ 5000 K and metallicities -1.5
$<$ [Fe/H] $<$ 0.5, making this sample ideal to compare with our bulge
sample. Spectra were taken with different spectrographs,
MIKE (R$\sim$65,000) mounted on the Clay 6.5m Magellan telescope, the 2dcoude
(R$\sim$60,000) spectrograph on the 2.7m Harlan J. Smith telescope at McDonald
Observatory, HIRES (R$\sim$100,000) on the Keck I 10m telescope and complemented
with spectra from the ELODIE archive
(R$\sim$42,000). Population membership and data reduction is described in AB10.
Population membership for the bulge sample is described in Z08.


\section{Stellar parameters}
For bulge stars we have  adopted the same stellar parameters, 
effective temperature (T$_{eff}$), surface gravity ($\log g$), microturbulence 
velocity ($\xi$) and metallicity [Fe/H] presented
in  Z08 and we refer the reader  to that  work  for a  detailed
description  of the  procedure.  In  order to  perform  a differential
analysis when  comparing bulge  and disk stars,  we followed  the same 
procedure from  Z08  to determine  parameters of the disk
sample. We have used the same iron linelist from Z08 based on calibrated $\log gf$ values to match the observed spectra of the metal-rich giant $\mu$-Leonis. Only small difference is that as 
first input value for T$_{\rm eff}$ we used photometric values
obtained from the calibration provided by \citet{alonso99} as opposed to the
\citet{ramirez05} calibration adopted by Z08. T$_{\rm eff}$ values were then
refined spectroscopically by imposing excitation  equilibrium between
FeI lines therefore removing any dependance on the adopted calibration. 
Microturbulence velocity $\xi$ was calculated by requiring
the  same [Fe/H]  for all  equivalent  widths of  FeI lines.   Surface
gravity was determined photometrically using the following equation:

\begin{footnotesize}
\begin{equation}
\log\left(g\right)=\log\left(g_{\odot}\right)+\log\left(\frac{M_*}{M_{\odot}}
\right)+0.4\left(M_{Bol,*}-M_{Bol,\odot}\right)+4\log\left(\frac{T_{eff,*}}{T_{eff,\odot}}\right)
\end{equation}
\end{footnotesize}

\noindent  where for  the Sun  we have  used the  same values  used by
Zoccali  et  al.   (2008): bolometric magnitude  $\rm  M_{Bol,\odot}$  =  4.72  mag, effective temperature $\rm
T_{eff,\odot}$ = 5770 K and surface gravity $\log\left(g_{\odot}\right)$ = 4.44 dex. We have
adopted   the   masses  from   AB10   which   are   based  on   Padova
isochrones.   Absolute  visual   magnitudes  were   obtained  assuming
distances from  Hipparcos parallaxes.  Finally,  bolometric magnitudes
were    calculated   by    applying   bolometric    corrections   from
\citet{alonso99}.
\noindent \\

\section{Abundance analysis}

Mg, Ca, Ti and Si abundances  were determined  by comparing the observed spectra with synthetic
ones created with MOOG \citep{sneden73}.  MOOG is  a FORTRAN code that
performs a spectral line analysis and spectrum synthesis assuming local thermal
equilibrium. MARCS model atmospheres \citep{gustafsson08} were used for our analysis. These models were created using the stellar parameters provided in Z08 for bulge stars, while for disk stars we used the stellar parameters determined as described in Section~3.
We have used the same atomic line list from the analysis of bulge stars in \citet[][hereafter L07]{lecureur07}. Unfortunately, the lower resolution of GIRAFFE compared to that of the UVES spectra of L07 prevents us to obtain accurate oxygen abundance measurements from the available OI line at 6300 \AA{}.

The broad ($\sim$5 \AA{}) autoionization Ca I line at 6318.1 \AA{} affects the continuum of the Mg triplet as discussed in L07. Therefore, Ca abundances were determined first in our procedure and then this value was given as an input to determine Mg abundances. The CN line list from the same work was adopted and calibrated carefully as CN lines are known to be important for cool giant stars. The TiO molecular line list \citep{plez98} was also
included. Reference solar abundances are from \citet{asplund09}.

\begin{table}
\begin{center}
\caption{Atomic linelist for Ca, Mg, Ti and Si used in this work. Also listed are the excitation potential ($\chi_{\rm ex}$) and oscillator strength ($\log gf$) for each analysed line.\label{lines}}
\begin{tabular}{c c c c c}
\\[3pt]
\hline
$\lambda$ (\AA) & Element & \large{$\chi_{\rm ex}$} & $\log gf$\\
\hline
6318.71 &  MgI  & 5.110 & -2.000\\
6319.23 &  MgI  & 5.110 & -2.240\\
6319.49 &  MgI  & 5.110 & -2.680\\
6166.43 &  CaI  & 2.521 & -1.142\\
6169.04 &  CaI  & 2.523 & -0.797\\
6169.56 &  CaI  & 2.526 & -0.478\\
6312.23 &  TiI  & 1.460 & -1.552\\
6142.49 &  SiI  & 5.620 & -1.500\\

\hline
\end{tabular}
\end{center}
\end{table}

\begin{figure}
\begin{center}
\includegraphics[scale=0.38,angle=0]{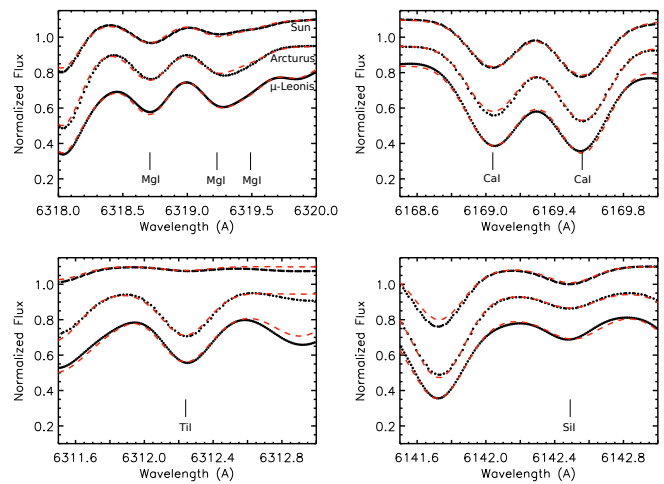}
\caption{Calibration of the synthesis regions for the Sun, Arcturus and $\mu$-Leonis. Observed spectra convolved to the GIRAFFE resolution are shown as black circles and the red solid line shows the synthetic spectra using the stellar parameters from L07 and abundances in Table~\ref{mu}.}
\label{muleo}
\end{center}
\end{figure}

Although this is a differential analysis, in the sense that we determine the stellar parameters and perform the spectral analysis in the exact same way for all stars, it is important to provide abundances of some reference targets using our procedure, therefore establishing our ``zero point'' in 
order to be able to contrast
our results with the values obtained in other works. For this reason we have also adopted the
calibration procedure based on reproducing the observed spectrum of the Sun, Arcturus
and $\mu$-Leonis. We have used the optical spectra of these stars that were analyzed in L07 for the same purpose. $\mu$-Leonis spectrum was taken at the Canada-France-Hawaii with the ESPanDOnS spectropolarimeter at a resolution of 80,000. The Arcturus spectrum with a resolution of 120,000 comes from the UVES database \citep{bagnulo03}. Finally for the Sun we used a UVES optical spectrum\footnote{(http://www.eso.org/observing/dfo/quality/UVES/-pipeline/solar\_spectrum.html)}. These spectra were convolved to the GIRAFFE resolution and compared with synthetic spectra in a region around the Mg, Ca, Ti, and Si lines analyzed in this work. Table~\ref{lines} shows the wavelength, excitation potential, and log gf of these lines. Model atmospheres were created adopting the same stellar parameters used in L07, [Fe/H]=0.3, T$_{\rm eff}$=4540, $\xi$=1.3, $\log g$=2.3 for $\mu$-Leonis, [Fe/H]=-0.52, T$_{\rm eff}$=4300, $\xi$=1.5, $\log g$=1.5 for Arcturus and T$_{\rm eff}$=5770, $\xi$=0.9, $\log g$=4.42 for the Sun. Figure~\ref{muleo} shows the comparison between the synthetic and observed spectra for these stars with the abundances listed in Table~\ref{mu}. For the Sun, we only required a 0.05 dex modification in Ca abundance with respect to our reference values of \citet{asplund09} also listed in Table~\ref{mu}.

\begin{table}
\begin{center}
\caption{Abundances of the Sun, Arcturus and $\mu$-Leonis, obtained by comparison to synthetic spectra, as well as reference solar abundances from \citet{asplund09}. The last two rows show the zero point abundances adopted by AB10 and our mean abundances measured for those same disk giants to which they refer.\label{mu}}
\begin{tabular}{c c c c c}
\\[3pt]
\hline
Stars & A(Ca) & A(Mg) & A(Ti) & A(Si)\\ 
\hline
Sun	     & 6.39  & 7.60 & 4.95 & 7.51\\
Sun \tiny{(Asplund et al)} & 6.34  & 7.60 & 4.95 & 7.51\\
Arcturus     & 5.93  & 7.30 & 4.55 & 7.17\\
$\mu$-Leonis & 6.65  & 8.02 & 5.31 & 7.85\\
\hline
AB10 ZP	     & 6.32  & 7.65 & 4.83 & 7.60\\
This work    & 6.49  & 7.65 & 5.09 & 7.62\\
\hline
\end{tabular}
\end{center}
\end{table}

\begin{figure}
\begin{center}
\includegraphics[scale=0.36,angle=0]{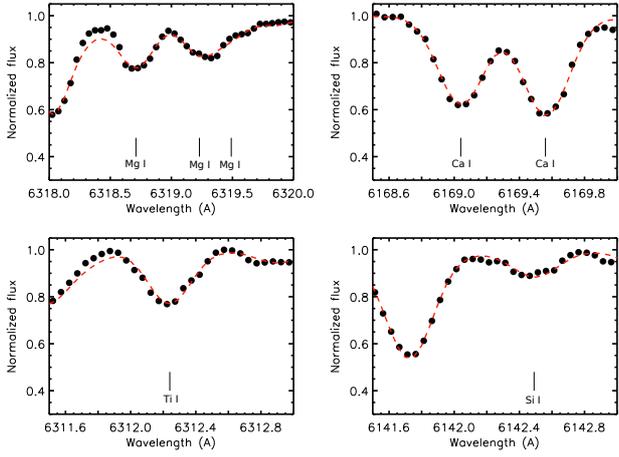}
\caption{Example of the final output of the synthesis procedure in order to derive the abundances of Mg (upper left), Ca (upper right), Ti (lower left) and Si (lower right). Observed spectrum for target star 212175c6 is plotted as black dots and the red solid line shows the synthetic spectrum using the best fitting abundance as obtained from our procedure.}
\label{synthesis}
\end{center}
\end{figure}

Abundances for our target stars were determined by an iterative process in which the synthetic spectrum is compared to the observed one modifying the abundance in each step of the process until reaching the best fitting value. Given the large number of stars in our sample, we have developed a code which calculates the best fitting abundance by $\chi^2$ minimization in a semi-automatic way. In the first step, radial velocities and a first guess for the FWHM of each star is determined using the DAOSPEC code \citep{stetson08}. The placement of the continuum is carried out manually for each star, by direct comparison between observed and synthetic spectrum in a region of 10 \AA{} around the line of interest. Additionally, inspection of nearby lines are used to modify the FWHM from the value obtained from DAOSPEC, when the latter clearly fails to reproduce the spectral features. Later on, the code automatically calculates the best fitting abundance by minimization of a $\chi^2$ value which is obtained in the following way:

\begin{equation}
\chi^{2}=\Sigma_{\lambda}(\mathrm{F}_{syn}(\lambda)-\mathrm{F}_{obs}(\lambda))^{2}\cdot \mathrm{W}(\lambda)
\end{equation}
where $F_{syn}(\lambda)$ is the flux of the synthesis at each wavelength, $F_{obs}(\lambda)$ is the observed flux previously normalized and W$(\lambda)$ is a weight value calculated as the difference between a normal synthesis and another without the line of the element to be fitted. This allows us to give more importance to the core of the line under analysis. To create this weight we have used $\mu$ Leonis parameters and abundances as a reference \citep{gratton+sneden90}. Figure~\ref{synthesis} shows an example of the final output from the code for each analyzed line. 

\section{Error analysis}

\subsection{Error due to stellar parameters uncertainties}
\begin{figure*}
\begin{center}
\includegraphics[scale=0.45,angle=180]{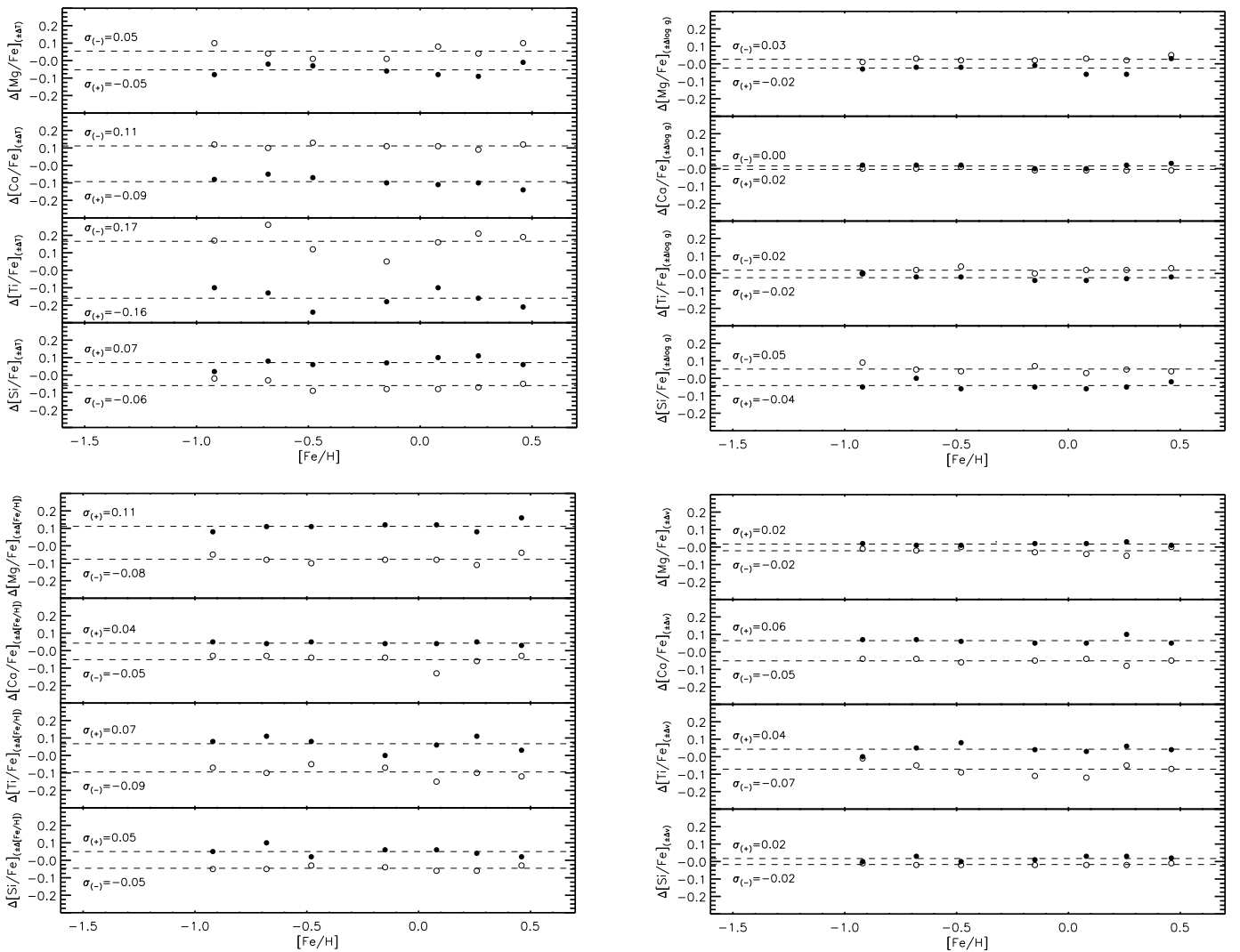}
\caption{Abundance errors associated with the uncertainties in effective temperature (upper left), surface gravity (upper right), metallicity (lower left) and microturbulence (lower right). A change of $\pm$200 K was applied to T$_{\rm eff}$, $\pm$0.1 dex to [Fe/H], $\pm$0.3 dex to $\log g$ and $\pm$0.2 dex to $\xi$. Model atmosphere were then created and abundances were re-determined. Filled circles show the change in abundance as $\Delta$[X/Fe]=[X/Fe]-[X/Fe]$_{\pm\Delta}$ when changes were added to the original parameters. Empty circles show the change in abundance when the respective value is subtracted. Dashed lines show the mean variations $\sigma_{(+)}$ and $\sigma_{(-)}$ when the uncertainty in the stellar parameter is added and subtracted respectively.}
\label{err_t}
\end{center}
\end{figure*}
We have carefully analyzed the effect of the different sources of error in our procedure. When obtaining stellar abundances the errors are commonly dominated by the uncertainties in the stellar parameters. Their influence can be determined by varying the adopted parameters by their respective uncertainty and then re-calculating the abundance value which best fits the observed spectra. We have carried out this analysis in a subsample of stars which covers the whole range of metallicities and effective temperatures of our sample. We have used the uncertainties in stellar parameters presented in Z08 of 200K for T$_{\rm eff}$, 0.3 dex in $\log g$, 0.2 dex in $\xi$ and 0.1 dex in the final [Fe/H] value. Figure~\ref{err_t} shows the variation in determined abundances for each element when varying the stellar parameters by their respective uncertainty. The final errors $\sigma_{T}$,$\sigma_{\rm [Fe/H]}$,$\sigma_{\log g}$,$\sigma_{\xi}$ associated to each parameter were calculated as the average between the mean changes in abundance when uncertainties were added ($\sigma_{(+)}$) and subtracted ($\sigma_{(-)}$).    

Figure~\ref{err_t} shows that errors in the abundance determination are strongly dominated by the uncertainties in the effective temperature determination. Given the adopted uncertainty of 200 K in temperature, Mg seems to be the least sensitive element with an associated error of $\sigma_{\Delta T}$=0.05 dex. Ca and Si show an associated error of 0.07 and 0.10 dex, respectively, and Ti has the largest dependence on temperature with an error value of $\sigma_{\Delta T}$=0.17 dex. From the lower left panel of Fig.~\ref{err_t} we see that the errors associated to the uncertainties in metallicity are quite similar among the elements with $\sigma_{[Fe/H]}$ values ranging between 0.05 and 0.10 dex for an adopted 0.1 dex error in [Fe/H]. Contrary to the sensitivity of the other elements, Mg is more affected by changes in metallicity than in temperature. The reason for such a dependence is most likely given by the blend with the strong FeI line at the red side of the triplet. Additionally Fig.~\ref{err_t} shows that the errors arising from uncertainties in gravity, $\sigma_{\log g}$, and microturbulence, $\sigma_{\xi}$, are small. However, Si still shows a 0.05 dex variation given an uncertainty of 0.3 dex in gravity. This is an important point to notice given that in our analysis we assume a distance of 8 kpc for all bulge stars, neglecting the bulge distance spread. If such stars are actually at 6 kpc, the bolometric magnitude will be underestimated by 0.65 mag. Such an error will change the photometric gravity by 0.25 dex.
Mg abundances are almost completely in-sensitive to changes in both $\log g$ and $\xi$ while Ca and Ti show a dependence of 0.06 dex on a microturbulence change of 0.2 dex.
        
\subsection{Errors from the spectral synthesis}

\begin{figure}
\begin{center}
\includegraphics[scale=0.45,angle=90]{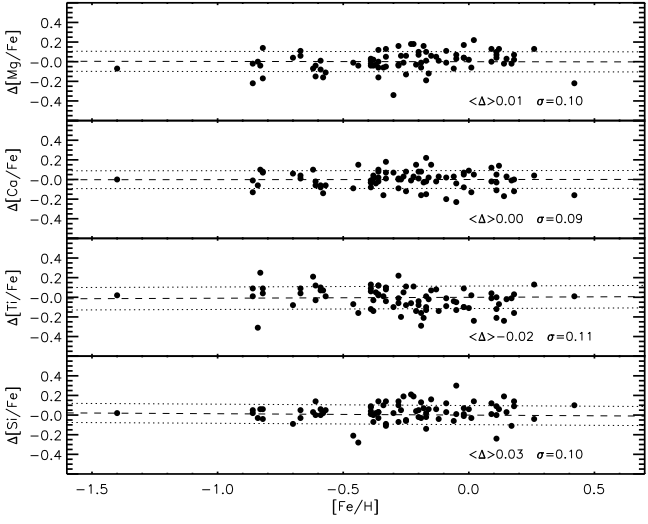}
\caption{Difference between the abundances measured using the same lines for stars observed twice in the field at b$=-12$. The dashed line shows the mean difference between both measurements and the dotted lines the 1$\sigma$ around the mean.}
\label{blanco_diff}
\end{center}
\end{figure}

As described in Z08, stars in the field at b$=-12$ were observed twice due to a mistake in the fiber allocation. For this reason we only have about half the number of stars than in the other fields. However, we can use this repeated observation to carry out the same procedure to re-derive the abundances for all the elements analyzed in our work. In this way, we can estimate the influence of the steps in our analysis which are done by \textit{eye} such as continuum placement and FWHM adjustment used for the synthesis.  Additionally this will also provide an estimate for the errors associated to different S/N among stars in the same field. Figure~\ref{blanco_diff}  shows the difference between the two independent analysis. The mean value of the difference for all elements ranges between 0 and 0.02 dex, which shows the absence of systematics in our procedure. The scatter is quite constant among the elements up to a value of 0.1 dex. Since the stellar parameters were not re-calculated, this scatter is not produced by their uncertainties but from the procedure to measure the abundances. We can therefore assume a 0.1 dex error from this source, which added to the errors from the stellar parameters gives an estimation for the final errors in our measured abundances following the relation:

\begin{equation}
\sigma^{2}_{\mathrm{[X/Fe]}}=\sigma^{2}_{\Delta \mathrm{T}}+\sigma^{2}_{\Delta \mathrm{[Fe/H]}}+\sigma^{2}_{\Delta \log g}+\sigma^{2}_{\Delta \xi}+\sigma^{2}_{syn}
\end{equation}

Therefore, the final values of the estimated error in our analysis are 0.15 dex for [Mg/Fe], 0.16 dex for 
[Ca/Fe], 0.22 dex for [Ti/Fe] and 0.14 dex for [Si/Fe].

\subsection{The role of spectral resolution}

It is also necessary to consider the effect of using lower resolution (R$\sim$22,500) spectra from GIRAFFE in comparison to the high resolution of the disk sample of R$>$45,000. In order to address this question we have re-calculated all the abundances using high resolution spectra for a subsample of stars in each field that was also observed with UVES(R$\sim$45,000). The abundances were calculated following the same procedure as for the GIRAFFE sample.
\begin{figure}
\begin{center}
\includegraphics[scale=0.45,angle=90]{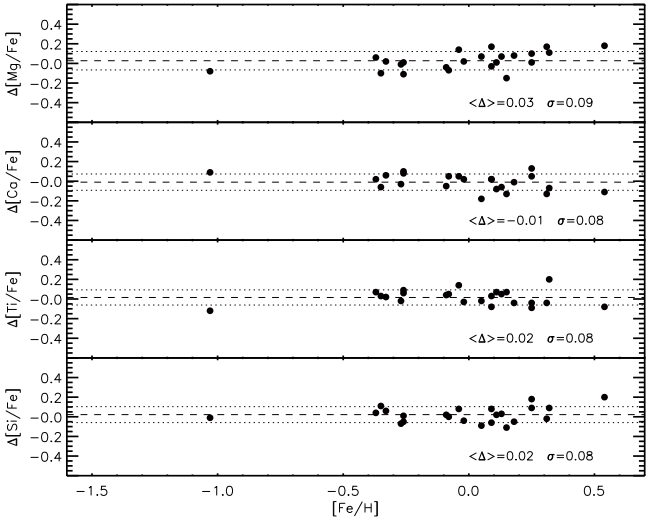}
\caption{$\Delta$[$\alpha$/Fe]=[$\alpha$/Fe]$_{UVES}$-[$\alpha$/Fe]$_{GIRAFFE}$ for our bulge sample plotted as black filled circles. The dashed line indicates the mean value of the difference and the dotted lines show the 1$\sigma$ range around the mean difference for each element.}
\label{uves}
\end{center}
\end{figure}

Figure~\ref{uves} shows the comparison between the abundances for each star in all analyzed elements. No systematic offsets are observed in our analysis from the change in resolution. This is particularly important to notice since it could be a source of systematic shifts in the trends when comparing bulge and disk.

\begin{figure}
\begin{center}
\includegraphics[scale=0.45,angle=90]{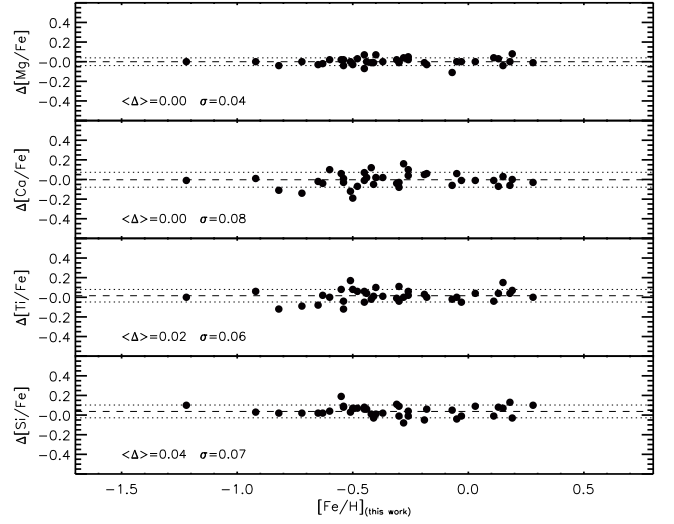}
\caption{$\Delta$[$\alpha$/Fe]=[$\alpha$/Fe]$_{HR}$-[$\alpha$/Fe]$_{LR}$ for the disk sample plotted as black filled circles where the low resolution (LR) spectra were obtained by degrading the high resolution (HR) to the resolution of our GIRAFFE sample. The dashed line indicates the mean value of the difference and the dotted lines show the 1$\sigma$ range for each element.}
\label{resolution}
\end{center}
\end{figure}

In order to further check the existence of any systematics we have also redetermined the abundances in the disk sample after degrading the resolution of the spectra to the one of GIRAFFE. Figure~\ref{resolution} confirms that no differences are observed from a change in resolution. Only Si shows a mean difference of 0.04 dex which is negligible within the errors on the abundance itself.

Additionally we have re-measured the abundances in our calibrator stars Arcturus, $\mu$-Leonis and the Sun, using the spectra with the original high resolution. The results were compared with our abundance measurements from convolved spectra (Table~\ref{mu} and Fig.~\ref{muleo}) and no differences were found. We therefore conclude that our comparison of bulge stars based on lower resolution to those of the disk is valid and no systematic shifts should be applied to our results.

\section{Results}

\subsection{Thick and thin disk sample}

\begin{figure}
\begin{center}
\includegraphics[scale=0.55,angle=90]{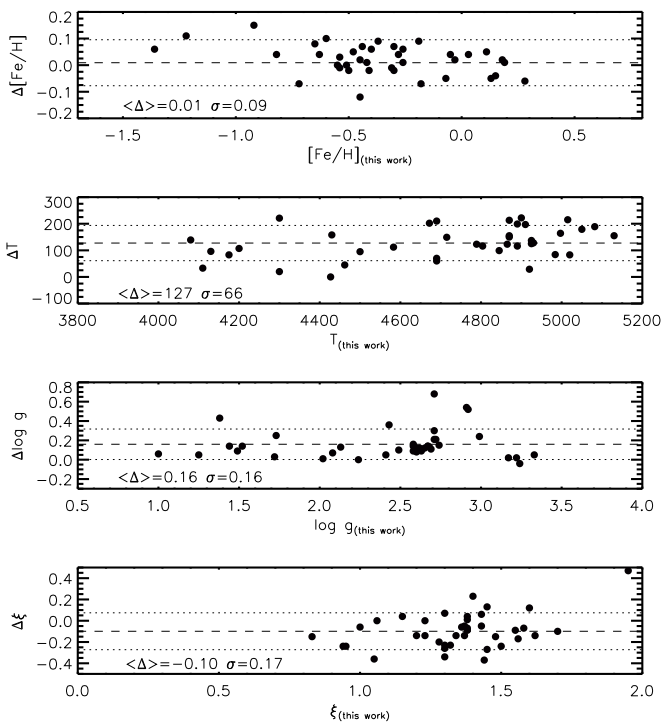}
\caption{Difference between the stellar parameters obtained in this work and the values presented in AB10 calculated as $\Delta$X=X$_{\rm this work}$-X$_{\rm AB10}$. The dashed line indicates the mean value of the difference and the dotted lines show the 1$\sigma$ range for each element.}
\label{diff_pars}
\end{center}
\end{figure}

In order to characterize the bulge alpha abundances and compare them to other galactic components we have checked carefully the homogeneity of the procedure. Stellar parameters for the disk sample were calculated following the same procedure described in Z08. A comparison between the stellar parameters obtained using our procedure and the values presented in AB10 is shown in Fig.~\ref{diff_pars}. We systematically find a larger effective temperature with a mean difference of 130 K between the two methods. Also, $\log g$ values are systematically larger in our analysis with respect to AB10. These differences between our studies show once more the importance of a homogeneous analysis in which parameters for all samples are on the same scale.        

\begin{figure}
\begin{center}
\includegraphics[scale=0.45,angle=90]{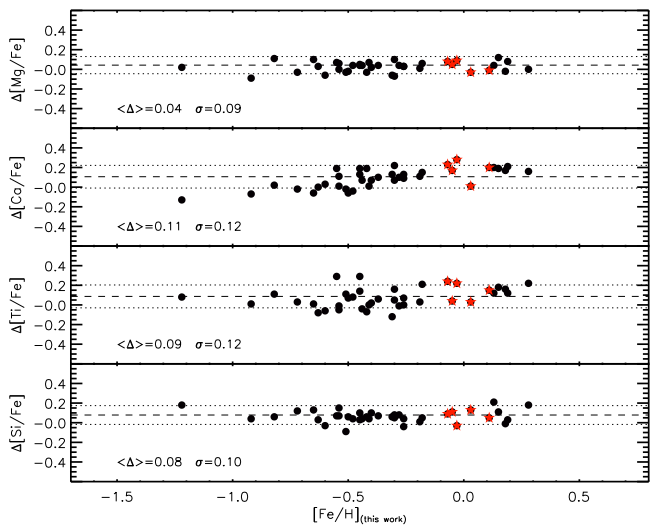}
\caption{Difference between the [$\alpha$/Fe] values for Mg, Ca, Ti and Si obtained in this work and the values presented in AB10 calculated as $\Delta$[$\alpha$/Fe]=[$\alpha$/Fe]$_{\rm this work}$-[$\alpha$/Fe]$_{\rm AB10}$.  The dashed line indicates the mean value of the difference and the dotted lines show the 1$\sigma$ range for each element. In particular, disk giants used by AB10 to calculate their internal zero points are shown as red filled stars.}
\label{diff_alan}
\end{center}
\end{figure}

Using our determination of the stellar parameters, we have measured the abundances for Mg, Ca, Ti and Si as described in section 3. Figure~\ref{diff_alan} shows the difference between the abundances for the disk sample derived in this work and the values presented in AB10. Mg abundances appear to be on the same scale however abundances of Ca, Ti show larger differences. 

AB10 uses the abundances of a subsample of thin disk stars, with metallicity near to the solar value, as internal zero-points. For the same stars we find larger Ca and Ti abundances as shown in Table~\ref{mu}. However we remain with our zero-points based on the solar abundances of \citet{asplund09}. Table~\ref{mu} should be used to compare our results either with models scaled to solar abundances or other works on different scales. Within our log-gf and adopted zero points, the homogeneity of our analysis remains unchanged.

\subsection{Alpha elements across the bulge, thick and thin disk}

\begin{figure}
\begin{center}
\includegraphics[scale=0.45,angle=90]{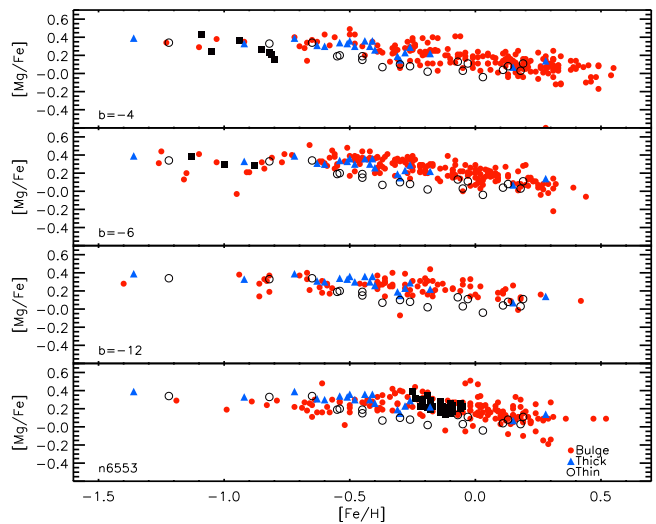}
\caption{[Mg/Fe] abundances in 4 fields of the
bulge shown as red filled circles. Bulge globular cluster members are shown as black filled squares. [Mg/Fe] abundances for the thick disk stars are shown as blue filled triangles and as empty black circles for the thin disk stars.}
\label{mg}
\end{center}
\end{figure}

\begin{figure}
\begin{center}
\includegraphics[scale=0.45,angle=90]{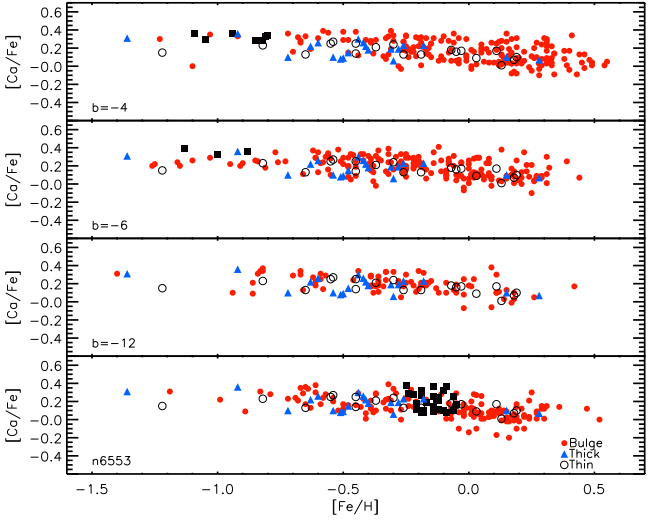}
\caption{[Ca/Fe] abundances in 4 fields of the bulge. Symbols are as in Fig.~\ref{mg}.}
\label{ca}
\end{center}
\end{figure}

\begin{figure}
\begin{center}
\includegraphics[scale=0.45,angle=90]{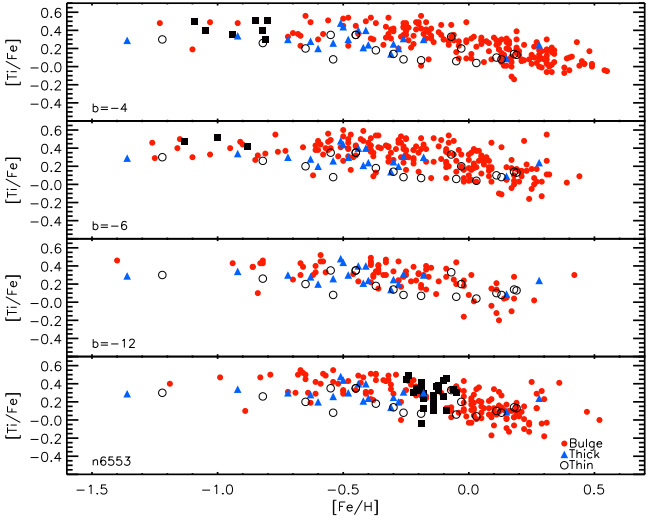}
\caption{[Ti/Fe] abundances in 4 fields of the bulge. Symbols are as in Fig.~\ref{mg}.}
\label{ti}
\end{center}
\end{figure}

\begin{figure}
\begin{center}
\includegraphics[scale=0.45,angle=90]{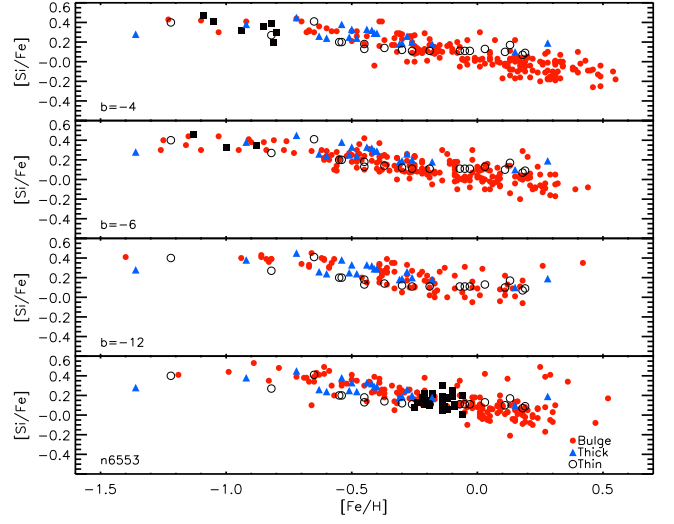}
\caption{[Si/Fe] abundances in 4 fields of the bulge. Symbols are as in Fig.~\ref{mg}.}
\label{si}
\end{center}
\end{figure}

The final [$\alpha$/Fe] trends for Mg, Ca, Ti, and Si are shown in Figure~\ref{mg} to \ref{si} and listed in Table~\ref{alphastab}. The general alpha enhancement of the bulge stars is observed clearly in all four elements. The alpha enhancement is also seen in our globular cluster members and their mean abundances do not differ from those of field stars at their given metallicity. The scatter in the observed trends is different: Mg abundances show much smaller scatter than Ca, Ti and Si abundances. In particular at low metallicities, where there is almost no dependence on [Fe/H], there is a $\sim$0.09 dex scatter observed for [Mg/Fe] and $\sim$0.13 dex for the other elements. As shown in section 5, the dependence of the measured abundance on the stellar parameter uncertainties also varies among elements and the observed scatter in abundances is within the estimated errors. Therefore, the different amount of scatter is largely due to the uncertainties in the parameters. The similarity and overabundance in the bulge and thick disk stars appears to vary among the elements and therefore trends should be compared separately. 

\begin{table}
\begin{center}
\caption{Mg, Ca, Ti and Si abundances for stars in the four bulge fields. Metallicities are those presented in Z08 and listed here for reference. The full table is available in electronic form.\label{alphastab}}
\begin{tabular}{c c c c c c}
\\[3pt]
\hline
star ID & [Fe/H] & [Mg/Fe] & [Ca/Fe] & [Ti/Fe] & [Si/Fe]\\ 
\hline
423342 & +0.46 &  -0.04 &  +0.13 &   +0.03   &  -0.08\\
423323 & -0.48 &  +0.43 &  +0.15 &   +0.23   &  +0.26\\
412779 & -0.37 &  +0.23 &  +0.29 &   +0.48   &  +0.22\\
423359 & -1.23 &  +0.34 &  +0.30 &   +0.48   &  +0.43\\
...... & ..... &  ..... &  ..... &   .....   &  .....\\
\hline
\end{tabular}
\end{center}
\end{table}

Mg is the element which shows the best defined trend among the elements analyzed here. This element shows the least dependency on the stellar parameters and can be considered as the most reliable of the alpha elements for chemical enrichment studies as, together with oxygen, it is expected to be enriched exclusively by SNII explosions. The mean abundance of the bulge at [Fe/H]$<-0.5$ is [Mg/Fe]$=+0.31$ which only differs by 0.03 dex from the mean thick disk at the same metallicity range with the bulge being slightly more enhanced than the thick disk. Such a behavior remains in the intermediate metallicity range of our sample ($-0.5<\mathrm{[Fe/H]}<-0.2$) where the bulge shows a mean [Mg/Fe] of 0.27 and is still only 0.03 dex more enhanced that the thick disk. The differences we have found between these populations are much smaller than those found by L07, in which the bulge is up to 0.2 dex more enhanced than the sample of thick disk dwarfs from the literature. Further comparison at higher metallicities are difficult as separation between thin and thick disk becomes difficult. All elements show that, at the high metallicity range ([Fe/H]$>0$) both the bulge and two of the high metallicity stars classified as thick disk by AB10 have lower ratios, as low as those of the thin disk.  

On the other hand, [Ca/Fe] trends show a larger scatter which complicates the comparison, particularly at high metallicities. However, at lower metallicities, where actually the thick disk is well defined, differences in [Ca/Fe] between the mean of both populations are higher showing the bulge 0.09 dex more enhanced than the thick disk at [Fe/H]$<-0.5$ and 0.04 dex at higher metallicities. Worth notice as well is that [Ca/Fe] shows a shallower decline with metallicity than Mg in good agreement with model predictions where Ca contribution is also expected from SNIa explosions.

Results for [Ti/Fe] are similar to those observed in Ca, showing trends with a shallower decline with metallicity but also a larger scatter. The mean abundance is [Ti/Fe]=0.40 and shows a difference with the thick disk of 0.10 dex from [Fe/H]$<$-0.5 up to intermediate metallicities. Only above solar metallicities the mean Ti abundances of bulge and thick disk are the same. However these differences are within the scatter of the Ti abundances in the bulge which is consistent with the larger error dependence on temperature. Moreover, the [Ti/Fe] ratios in the thick disk are actually within the lower envelope of those of the bulge.

Finally in the case of [Si/Fe], trends for the disk and the bulge are almost undistinguishable as in the case of Mg, with the bulge showing a mean [Si/Fe]=0.35 at [Fe/H]$<-0.5$ and differences lower than 0.03 dex with the thick disk at all metallicities.  
   
\begin{figure}
\begin{center}
\includegraphics[scale=0.45,angle=0]{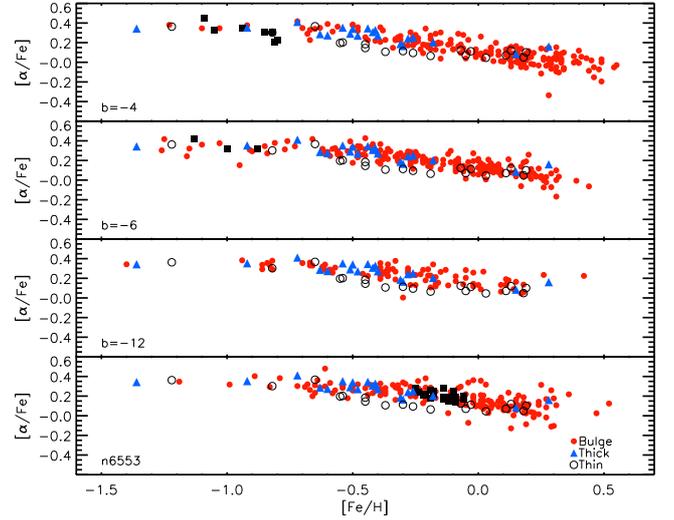}
\caption{[$\alpha$/Fe] abundances in 4 fields of the bulge calculated as the average between Ca, Mg, Ti and Si abundances. Symbols are as in Fig.~\ref{mg}.}
\label{alphas}
\end{center}
\end{figure}

Figure~\ref{alphas} shows the average $\alpha$ element abundances. Trends for the bulge and both thin and thick disk are very well defined and we can also use them to compare the population properties between our different fields along the minor axis and with the field at a larger galactic longitude of l=5$^\circ$. A discussion regarding such a comparison is addressed in section 6.3.

\begin{figure*}
\begin{center}
\includegraphics[scale=0.45,angle=180]{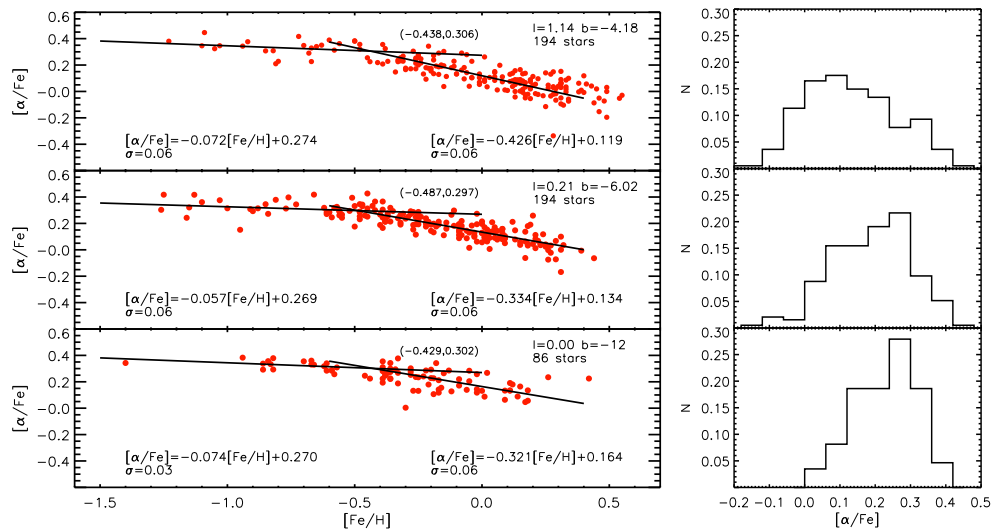}
\caption{The left panels show [$\alpha$/Fe] trends as a function of [Fe/H] in 3 bulge fields located along the minor axis. Best fit trends are shown for both [Fe/H] ranges, a metal-poor between $-1.2$ and $-0.5$ dex and a metal-rich between $-0.3$ and $0.2$ dex, as well as the location of the knee in all fields.The right panels show the [$\alpha$/Fe] distribution for each field.}
\label{breakminor}
\end{center}
\end{figure*}

\begin{figure*}
\begin{center}
\includegraphics[scale=0.40,angle=0]{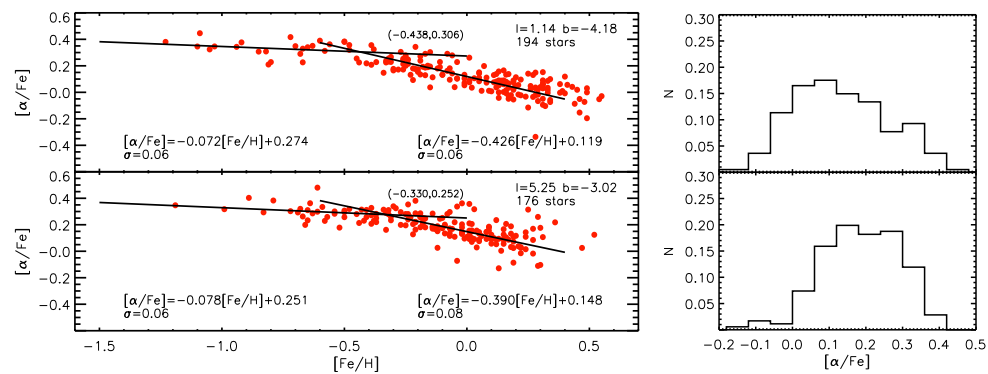}
\caption{Same as in Figure~\ref{breakminor} for 2 bulge fields along the major axis.}
\label{breakmayor}
\end{center}
\end{figure*}

In order to generally compare abundances between bulge and disk, we have followed the procedure adopted in AB10 and fit two independent slopes for two different metallicity regimes. However, instead of fixing arbitrarily the point where the knee is produced to a [Fe/H] value of -0.5, as done in AB10, we have produced a linear fit to a metal-poor region between $-1.2$ and $-0.5$ dex and to a metal-rich region between $-0.3$ and $0.2$ dex. These regions where selected in order to i) consider metallicity regions in which the trends are well defined and ii) avoid taking into account too metal-rich stars in which the uncertainties are larger. In this way, we are now able to provide a value for the metallicity at which the knee is actually produced, calculated as the intersection of the linear fits of both regions. The best fitting relations for both metallicity ranges in all the bulge fields and the thick disk are shown in Table~\ref{linear} and plotted in Figures~\ref{breakminor} and \ref{breakmayor}.

\begin{table}
\newcommand\T{\rule{0pt}{2.6ex}}
\newcommand\B{\rule[-1.2ex]{0pt}{0pt}}
\begin{center}
\addtolength{\tabcolsep}{-3pt}
\caption{Best fit parameters for linear (a+m[Fe/H]) trends for [$<$Mg,Ca,Ti,Si$>$/Fe] abundances in the bulge and thick disk.\label{linear}}
\tiny
\begin{tabular}{l c c c c}
\\[3pt]
\hline
\hline
\T \B & \multicolumn{2}{c}{$-1.2<\mathrm{[Fe/H]}<-0.5$} & \multicolumn{2}{c}{$-0.3<\mathrm{[Fe/H]}<0.2$}\\
\hline
\T \B Field & m & a & m & a \\ 
\hline

\T Baade's Window & -0.07$\pm$0.07 & 0.27$\pm$0.06 & -0.43$\pm$0.04 & 0.119$\pm$0.006\\
$b=-6^{o}$ & -0.06$\pm$0.04 & 0.27$\pm$0.04 & -0.33$\pm$0.04 & 0.134$\pm$0.006\\
Blanco field & -0.07$\pm$0.04 & 0.27$\pm$0.03 & -0.32$\pm$0.06 & 0.16$\pm$0.01\\
NGC 6553 & -0.08$\pm$0.09 & 0.25$\pm$0.06 & -0.39$\pm$0.06 & 0.148$\pm$0.008\\
Thick disk & -0.04$\pm$0.06 & 0.30$\pm$0.05 & -0.37$\pm$0.08 & 0.13$\pm$0.03\\
\hline

\end{tabular}
\end{center}
\end{table}

As a general result, when considering the mean abundances of all the elements analyzed, differences between the bulge and the thick disk do not exceed 0.05 dex for all metallicities. Trends between both populations are therefore indistinguishable within the errors of our analysis. Additionally we see that the knee in the bulge in Baade's Window is located at a metallicity of [Fe/H]=-0.44 and differs by less than 0.1 dex among the other bulge fields. This knee is also present in the thick disk distribution at [Fe/H]=-0.51, only 0.07 dex different to the location for the observed bulge downtrend. Therefore, within the errors considered in our analysis as discussed in previous sections, no significant differences are observed between the bulge and the thick disk abundances for individual elements. The similarity is even stronger when all elements are compared in the mean.

\subsection{Is there an $\alpha$-element gradient in the bulge?}

The existence of a metallicity gradient in the bulge is considered a strong indication towards the classical bulge formation scenario and is not expected to be present in boxy or pseudo-bulges. 
On the other hand, if Galactic bulge is a mixture of two (or more) components, it could be expected that these populations should have gone through different formation histories. A component formed via mergers would show higher alpha elements than a structure dynamically formed from disk material which would have enough time for SNIa to explode and decrease the [$\alpha$/Fe] ratio up to Solar values. These components might have different relative contributions across the bulge and therefore their content of alpha elements might help to distinguish between them.

Figures~\ref{breakminor} and \ref{breakmayor} show that in the case of the Galactic bulge, the relations between [$\alpha$/Fe] and metallicity are undistinguishable among the different fields analyzed here. From the obtained relations for each field and in the two metallicity regimes that we have considered here, we can actually see no significant difference between the fields in terms of the IMF and the timescale given by the knee at which SNIa contribution becomes important. Differences between the fields are not larger than 0.02 dex at all metallicity regimes and the knee is observed at the same metallicity within 0.1 dex. This variation is in fact within our estimation for the uncertainties in [Fe/H]. Such small differences are fairly consistent with the errors considered in our abundance analysis and a striking similarity can be concluded both along the minor axis (Fig.~\ref{breakminor}) and the major axis (Fig.~\ref{breakmayor}). Also shown in Fig.~\ref{breakminor} and Fig.~\ref{breakmayor}, are the [$\alpha$/Fe] distributions for each of the fields. The behavior of such distributions are consistent with the metallicity gradient in the bulge in which the population of stars with higher metallicities ([Fe/H]$\sim$0) and lower $\alpha$ element abundances ([$\alpha$/Fe]$\sim$0) becomes smaller at higher distances from the galactic plane.
 
\subsection{Correlation with kinematics}

\begin{figure}
\begin{center}
\includegraphics[scale=0.50,angle=90]{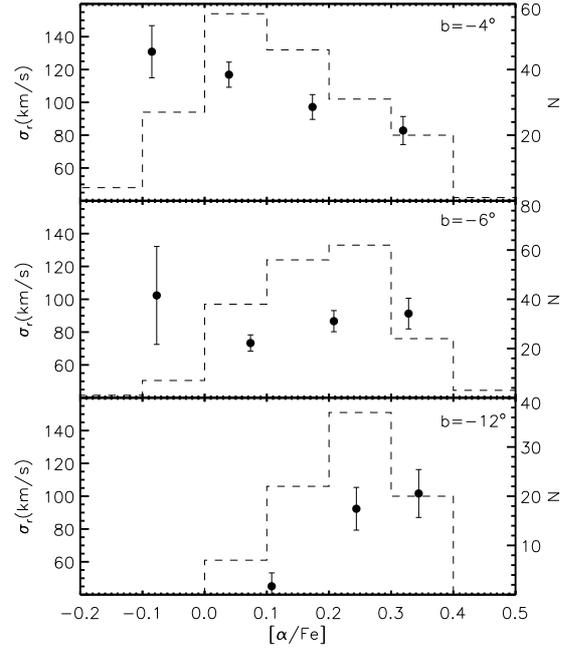}
\caption{Dispersion of radial velocity as a function of [$\alpha$/Fe], in bins of 0.15 dex, for the three fields along the minor axis (filled circles and left axis scale). Overploted as a dashed line is the [$\alpha$/Fe] distribution for each field, which scale is given on the right. Globular cluster members in the sample were removed for this particular analysis.}
\label{radial}
\end{center}
\end{figure}

Coupling the information from radial velocities,  proper motions and metallicity for the same bulge FLAMES sample that we present here, \citet{babusiaux10} demonstrate that metal-rich component shows disk/bar kinematics and a metal-poor component has kinematics of an old spheroid. The formation time-scale for the old (and metal-poor) spheroid component is much shorter, than that of the disk/bar component, and therefore it is expected that the former is alpha enhanced, while the latter has lower alpha element abundance. We investigate this by combining our alpha element abundances with radial velocity information from Z08.  Figure~\ref{radial} shows the radial velocity dispersion for different alpha element abundance bins. We observe that alpha-enhanced stars have a velocity dispersion of $\sim$90 km/s, which is approximately constant along the minor axis, while the stars with solar or slightly sub-solar $[\alpha / \mathrm{Fe}]$ abundances have larger velocity dispersion in the inner fields and decreasing velocity dispersion at higher latitude. In the b=-12$^\circ$ field the most alpha-poor component disappears altogether. The behaviour of alpha-enhanced stars mimics very closely the (lack of) trend shown by metal-poor component identified as having spheroid kinematics, while alpha-poor stars follow the trend of metal-rich component identified with disk/bar kinematics by \citet{babusiaux10}. While these trends are interesting, they need to be confirmed with a more complete coverage of the bulge.

\section{Discussion}

We have confirmed the chemical similarity between the alpha element abundances in bulge and 
thick disk within the errors associated to our homogeneous abundance analysis. An overabundance in alpha elements of [$\alpha$/Fe]=0.30 dex up to a metallicity of [Fe/H]$\sim$-0.5 is observed in both components as an indication of a short timescale of formation, with a chemical enrichment dominated by the contribution of core collapse supernovae. At this metallicity the observed downtrend in the bulge reflects the instance in which the Fe contribution from SNIa becomes important. In the bulge both the trend at low metallicities and the value at which the knee is produced do not vary along the minor axis for latitudes b$<$-4. This can be interpreted as a formation process of the dominating population of the bulge being fairly homogeneous, sharing a similar IMF and formation timescale. Figure~\ref{breakmayor} shows that these trends are indistinguishable also for two fields at different longitudes (l=1$^\circ$ and l=5$^\circ$) at a fixed latitude of b=-4$^\circ$, an indication that the homogeneity is present along the major axis as well.   
 
The chemical similarity of bulge stars and those of the solar neighborhood thick disk, at least for metallicities in which thick and thin disk can be clearly separated, is also remarkably intriguing. Furthermore, \citet{bensby10b} demonstrated for the first time that inner disk stars show both the dichotomy thin-thick disk and the chemical similarity with the bulge. Either the bulge and thick disk went through similar formation histories or the process which formed the bulge is directly linked to that of the thick disk. Unfortunately the formation history of the thick disk is currently a matter of debate. Therefore, as long as a complete understanding of the formation mechanism of the thick disk is missing, a correct interpretation for the bulge--thick disk connection will be hard to achieve. 
Particularly interesting, in the context of our results, are the scenarios in which both bulge and thick disk share a fast and early formation such as the scenario discussed in \citet{Bournaud09} \citep[see also][]{immeli04}. They show the properties obtained from simulations in which a thick disk and bulge are formed internally by instabilities of gas-rich clumpy disks in young galaxies. Such clumpy and turbulent rotating disks are in fact observed in galaxies around z$\sim$2 \citep{elmegreen08}. This shared formation mechanism could explain the observed chemical similarities between these components.

The main result our observations provide is that all fields in the bulge follow the same [$\alpha$/Fe] vs.\ 
[Fe/H] relation, and that metal-poor stars correspond to a population enhanced in $\alpha$ elements which shows homogeneous kinematics. These observational facts taken on their own are in accordance with the classical bulge formation through the fast dissipative collapse. This is in contrast with the recent literature claims of pure disk/bar origin of the Galactic bulge \citep{shen10}, which would result in shallower [$\alpha$/Fe] decline, more similar to that of the thin disk. 
The most metal-rich part of the population, that shows a downturn in [$\alpha$/Fe] has formed on a timescale long enough for SNIa to start contributing significantly with iron peak elements and shows kinematics corresponding to a disk/bar component according to Babusiaux et al. (2010). This component progressively dissapears with increasing distance from the plane.

Altogether, or results provide observational evidence for the coexistance of a classical bulge, within a bar-like component. Dual component bulges have also been observed in external galaxies in the work of \citet{peletier07} which shows an old elliptical-like component dominating the Bulge at higher distances from the plane, while a disk-like component is observed in the central regions. In particular, we highlight the recent results from \citet{williams11}. They show that the bulge of a pure disk galaxy, NGC 3390, rotates perfectly cilindrically, has chemical properties resembling those of the disk and shows no metallicity gradient. On the other hand, the boxy bulge of NGC 1381 shows more complicated rotation properties, is alpha enhanced with respect to the disk and shows a metallicity gradient. These properties for NGC 1381 strikingly resemble those presented here, in a star by star basis, for the Milky Way bulge. In term of simulations, the chemo-dynamical model of \citet{samland03} also predicts a composite bulge: an early collapse component and a metal-rich, alpha-poor population associated to a bar. 

Clearly, to settle the issue of the origin of the bulge and to situate it in the context of the formation of the other Galactic components, a detailed analysis of the kinematics and stellar abundances is necessary in a more complete coverage of the bulge and in the inner disk of the Galaxy.

\section{Conclusions}

We have analyzed the abundances of Mg,Ca,Si,Ti in four fields of the Galactic bulge. We have carried out a homogeneous analysis comparing [$\alpha$/Fe] ratios for a large sample of bulge stars (650) with those of giants belonging to the thin and thick disk in the Solar vicinity. We also compared bulge [$\alpha$/Fe] distributions across the different fields in our sample. Our conclusions can be summarized as follows:

\begin{enumerate}
\item We have confirmed the chemical similarity regarding alpha element abundances between the bulge and the thick disk when a homogeneous analysis is done and using only giant stars spectra. Both populations are enhanced compared to the thin disk and show a downtrend in [$\alpha$/Fe] starting at a metallicity of [Fe/H]$\sim$-0.4. At higher metallicities, close to solar, bulge [$\alpha$/Fe] ratios drop to values similar to those of the thin disk.  
\item The trends of [$\alpha$/Fe] ratios are indistinguishable among our three fields along the bulge minor axis. Analysis on a field at l=4$^\circ$, b=5$^\circ$ hints for a homogeneity also present along the major axis. The knee showing the downtrend in [$\alpha$/Fe] is observed in all fields at [Fe/H]$\sim$-0.4. A more complete mapping of different bulge regions is necessary to confirm the conclusions regarding the major axis lack of gradients as well to explore the [$\alpha$/Fe] distribution for inner bulge fields (b$>$-4).
\item The [$\alpha$/Fe] distributions in our different fields show that, while the population of metal-rich stars and low [$\alpha$/Fe] ratios observed at b=-4$^\circ$ seems to disappear at larger galactic latitudes, trends and distributions at low metallicities ([Fe/H]$<$-0.2), as well as velocity dispersion, remain unchanged at different latitudes. Such results are consistent with recent evidences for a dual nature of the bulge.  
\end{enumerate}

Our results demand future work to be based on a more complete coverage of the bulge. A better mapping of chemical and kinematical signatures, coupled with the complete morphological characterization to be obtained from a multi-epoch photometric survey such as the VISTA Variables in the Via Lactea \citep[VVV;][]{vvv10}, could provide the final step towards the observational characterization of the bulge.


\begin{acknowledgements}
We thank the anonymous referee for constructive comments which improved our paper. We acknowledge E. Valenti and S. Lucatello for very useful suggestions and discussions, and M. Williams for fruitful discussions in the topic of boxy-bulges. MZ, and  OG acknowledge support by Proyecto Fondecyt Regular 1085278. DM acknowledge support by Proyecto Fondecyt Regular 1090213.
MZ and  DM are partly supported  by the BASAL  Center for Astrophysics
and Associated Technologies PFB-06, the FONDAP Center for Astrophysics
15010003 and the MIDEPLAN Milky Way Millennium Nucleus P07-021-F. 
AAB acknowledges financial support by FONDECYT project 3100013. 
BB acknowledges grants from CNPq and FAPESP.
\end{acknowledgements}
\renewcommand*{\bibfont}{\small}
\bibliographystyle{aa}

\bibliography{mybiblio}

\end{document}